\documentclass[]{aa}
\usepackage{psfig}
\begin{document}

\thesaurus{08(08.15.1)}

\title{The radial pulsation of AI~Aurigae}

\author{L.L. Kiss\inst{1} \and B.A. Skiff\inst{2}}

\institute{Department of Experimental Physics and Astronomical Observatory,
University of Szeged, Szeged, D\'om t\'er 9., H-6720 Hungary \and
Lowell Observatory, 1400 West Mars Hill Road, Flagstaff AZ 86001, USA}

\titlerunning{The radial pulsation of AI~Aurigae}
\authorrunning{Kiss \& Skiff}
\offprints{l.kiss@physx.u-szeged.hu}
\date{}

\maketitle
 
\begin{abstract}

We present an analysis of eleven years of Str\"omgren $by$ photometry
of the red semiregular variable star AI~Aurigae. An early period determination
of 63.9 days is confirmed by the long-term light curve
behaviour. The light curve shows semi-regular changes
with a mean period of 65 days reaching an amplitude of 0\fm6 in
some cycles. The $b-y$ colour changes perfectly parallel
the V light curve, suggesting radial oscillation to be the
main reason for the observed variations. We estimate the main
characteristics of the star (mass, radius, effective temperature)
that suggest radial pulsation in fundamental or first overtone mode.

 
\end{abstract}

\section{Introduction}

The light variation of semiregular variables (SR) of subtype
SRa and SRb is generally associated with the pulsation of these
low and intermediate mass red giants located on the asymptotic giant branch
(AGB) of the Hertzsprung--Russell diagram. Since the characteristic
timescale of the variations is between 20 and 2000 days (usually
hundreds of days), semiregulars are typical targets of amateur
observers estimating visual brightnesses. There are
very few high-precision light, colour, and/or radial-velocity
measurement covering many cycles that confirm the pulsational
origin of the variations. Very recently, Lebzelter and his
co-workers (Lebzelter 1999ab, Lebzelter et al. 1999) presented
high-precision photometric and infrared radial-velocity data for a
sample of bright SR stars. Their results both illustrate the ability
of automatic telescopes to monitor this type of variable stars and support
the assumption of pulsation (see also Percy et al. 1996 on involving
amateur photoelectric observers into regular observations of small-amplitude
red variables, SARVs).

However, one cannot ad hoc accept pulsation to be the main reason
of variability. As has been noted by Percy \& Polano (1998),
virtually every M giant is variable in brightness. This has been
also supported by the large-scale photometric surveys (Hipparcos,
MACHO, OGLE) leading to the discovery of thousands of M-type variables.
It has been pointed out by, e.g. Lebzelter et al. (2000) and
Kiss et al. (2000), that specifying the physical reason responsible
for the variation requires simultaneous light, colour, and
radial-velocity data, as even consecutive cycles may show significant
deviations. This kind of long-term observation is crucial
when studying the possible mechanisms affecting the light changes
of SR variables. Besides pulsation, other contributors to
variability cannot be readily excluded, such as time dependent
surface inhomogeneities due to large convective cells (Schwarzschild
1975, Lebzelter et al. 2000), ellipsoidal deformation due
to the presence of a close companion, or spots on a relatively
rapidly-rotating star.
Most recently, Koen \& Laney (2000) gave a detailed
list of considerations concerning the possible reasons of variability
and noted, that there is still no definite conclusion on this issue.
Until the application of these mechanisms is better
understood, each case must be studied separately.

AI~Aurigae (HD 259342, $9\fm1 < V < 9\fm8$, $P=63\fd9$,
spectral type M5III - Chuadze 1973) is a poorly-studied red semiregular
(subtype SRa) variable star. In the General Catalogue of
Variable Stars (GCVS -- Kholopov et al. 1985-1988)
a period of 63\fd9 is listed with an early epoch of maximum
(JD 2426029, due to Beyer (1937)). Since then there has been no other
photometric measurement published for this star. Due to the relatively small
range of light changes, it is also neglected by the international
organizations of amateur variable star observers (American Assocation
of Variable Star Observers -- AAVSO, Association Francaise des
Observateurs d'Etoiles Variables -- AFOEV and Variable Star Observers'
League in Japan -- VSOLJ). The AAVSO International Database does not contain
this star, AFOEV has 16 points which are wrongly identified with
AI~Aur (they are 2 mag brighter), and VSOLJ collected only 30 individual
estimates between 1986 and 1999. Similar neglect is present concerning
other observations: there was only one radial velocity monitoring
($\langle v_r \rangle =+63$ km~s$^{-1}$, Feast et al. 1972),
while Dickinson \& Dinger (1982) listed AI~Aur among the negative
detections during their H$_2$O survey. There is no
metallicity determination in the literature.
Its infrared colours (see later) place the star among the ``blue''
SRs (Kerschbaum \& Hron 1992) and the presently available data
suggest the star to be a regular member of this group.

The main aim of this paper is to present a continuous photometric
monitoring of AI~Aur between 1985 and 1996 which reveals the
associated colour variations in addition to obtaining an accurate
light curve.
(We note, that the presented observations were acquired during a
large-scale photometric survey of solar type stars (Lockwood et al. 1997),
without any specific reason. Therefore, this study is a
by-product of that long-term observing programme.)
The observed behaviour
of AI~Aur is very likely to be mainly due to radial
pulsation, most probably in fundamental or low-degree overtone mode.
The next section deals with the data aquisition, while the light- and
colour-curves are discussed in Sect.\ 3 together with the
physical parameters of the star. A summary of the conclusions
is given in Sect.\ 4.

\section{Observations}

\begin{table*}
\begin{center}
\caption{Str\"omgren photometry of AI~Aurigae (MJD=JD$-$2400000)}
\begin{tabular} {|lll|lll|lll|lll|}
\hline
MJD       & $V$   &  $b-y$ &  MJD       & $V$   &  $b-y$ &  MJD       & $V$   &  $b-y$ & MJD       & $V$   &  $b-y$ \\
\hline
46351.000 &  9.182 & 1.348 &   46765.900 &   9.346 & 1.330 &  47529.783 &   9.451 & 1.357 &   48991.754 &   9.477 & 1.376\\
46352.996 &  9.192 & 1.353 &   46768.933 &   9.354 & 1.336 &  47535.796 &   9.384 & 1.348 &   49016.708 &   9.659 & 1.402\\
46353.996 &  9.205 & 1.348 &   46777.838 &   9.367 & 1.360 &  47541.771 &   9.261 & 1.347 &   49020.617 &   9.708 & 1.427\\
46357.987 &  9.259 & 1.352 &   46806.787 &   9.159 & 1.341 &  47542.725 &   9.261 & 1.350 &   49051.654 &   9.435 & 1.394\\
46362.975 &  9.357 & 1.368 &   46807.771 &   9.168 & 1.329 &  47544.725 &   9.248 & 1.345 &   49054.658 &   9.396 & 1.406\\
46375.958 &  9.609 & 1.413 &   46827.758 &   9.463 & 1.389 &  47578.683 &   9.389 & 1.382 &   49058.675 &   9.373 & 1.385\\
46376.954 &  9.615 & 1.401 &   46828.763 &   9.450 & 1.398 &  47579.679 &   9.407 & 1.379 &   49325.792 &   9.474 & 1.382\\
46385.942 &  9.693 & 1.421 &   46831.833 &   9.447 & 1.396 &  47585.704 &   9.455 & 1.389 &   49326.829 &   9.463 & 1.389\\
46388.917 &  9.714 & 1.422 &   46832.750 &   9.442 & 1.385 &  47586.658 &   9.471 & 1.392 &   49329.821 &   9.425 & 1.403\\
46389.921 &  9.717 & 1.440 &   46848.721 &   9.332 & 1.352 &  47592.658 &   9.536 & 1.414 &   49330.871 &   9.407 & 1.389\\
46403.917 &  9.457 & 1.387 &   46853.692 &   9.275 & 1.338 &  47599.671 &   9.522 & 1.391 &   49371.754 &   9.363 & 1.378\\
46420.813 &  9.236 & 1.361 &   46857.658 &   9.217 & 1.319 &  47606.671 &   9.423 & 1.385 &   49373.775 &   9.358 & 1.385\\
46421.817 &  9.240 & 1.359 &   46858.683 &   9.198 & 1.325 &  47618.633 &   9.232 & 1.356 &   49382.812 &   9.326 & 1.354\\
46431.875 &  9.297 & 1.373 &   46862.700 &   9.164 & 1.314 &  47623.633 &   9.195 & 1.342 &   49384.687 &   9.311 & 1.371\\
46449.850 &  9.547 & 1.399 &   46867.650 &   9.127 & 1.321 &  47624.625 &   9.196 & 1.334 &   49395.692 &   9.398 & 1.382\\
46450.742 &  9.561 & 1.420 &   46880.638 &   9.358 & 1.359 &  47626.625 &   9.194 & 1.347 &   49398.725 &   9.467 & 1.436\\
46457.783 &  9.563 & 1.373 &   46881.658 &   9.387 & 1.336 &  47640.638 &   9.430 & 1.392 &   49439.617 &   9.435 & 1.362\\
46463.767 &  9.483 & 1.396 &   46885.675 &   9.435 & 1.340 &  47826.929 &   9.181 & 1.321 &   49447.642 &   9.378 & 1.383\\
46483.675 &  9.200 & 1.356 &   46898.629 &   9.408 & 1.333 &  47843.900 &   9.357 & 1.324 &   49633.958 &   9.200 & 1.349\\
46484.662 &  9.535 & 1.380 &   46899.629 &   9.419 & 1.310 &  47880.813 &   9.375 & 1.328 &   49670.883 &   9.617 & 1.388\\
46486.700 &  9.206 & 1.364 &   47103.987 &   9.305 & 1.358 &  47896.683 &   9.328 & 1.343 &   49684.858 &   9.678 & 1.411\\
46487.679 &  9.205 & 1.359 &   47115.937 &   9.055 & 1.334 &  47915.692 &   9.445 & 1.353 &   49685.842 &   9.662 & 1.418\\
46490.700 &  9.239 & 1.357 &   47118.950 &   9.066 & 1.341 &  47919.754 &   9.519 & 1.373 &   49687.800 &   9.639 & 1.411\\
46496.667 &  9.355 & 1.373 &   47125.892 &   9.205 & 1.336 &  47921.779 &   9.539 & 1.369 &   49707.787 &   9.479 & 1.381\\
46510.671 &  9.509 & 1.391 &   47140.846 &   9.572 & 1.416 &  47925.683 &   9.530 & 1.379 &   49750.696 &   9.480 & 1.390\\
46511.671 &  9.529 & 1.373 &   47148.838 &   9.765 & 1.425 &  47932.683 &   9.481 & 1.366 &   49752.713 &   9.450 & 1.386\\
46512.662 &  9.535 & 1.380 &   47150.871 &   9.786 & 1.444 &  47969.667 &   9.090 & 1.323 &   49754.692 &   9.449 & 1.378\\
46516.650 &  9.557 & 1.382 &   47170.763 &   9.267 & 1.369 &  47972.650 &   9.195 & 1.336 &   49757.708 &   9.438 & 1.381\\
46530.650 &  9.430 & 1.360 &   47172.829 &   9.239 & 1.365 &  47974.671 &   9.266 & 1.366 &   49767.625 &   9.484 & 1.393\\
46534.638 &  9.345 & 1.362 &   47199.692 &   9.480 & 1.385 &  48218.854 &   9.249 & 1.374 &   49783.625 &   9.520 & 1.379\\
46539.629 &  9.220 & 1.363 &   47200.696 &   9.511 & 1.405 &  48294.700 &   9.495 & 1.396 &   49979.942 &   9.342 & 1.358\\
46544.650 &  9.133 & 1.335 &   47203.721 &   9.598 & 1.407 &  48296.646 &   9.501 & 1.394 &   50007.917 &   9.530 & 1.380\\
46547.654 &  9.108 & 1.341 &   47204.729 &   9.616 & 1.410 &  48322.608 &   9.677 & 1.400 &   50010.954 &   9.502 & 1.378\\
46550.638 &  9.113 & 1.325 &   47205.729 &   9.636 & 1.407 &  48597.858 &   9.440 & 1.368 &   50034.871 &   9.082 & 1.330\\
46678.979 &  9.280 & 1.370 &   47206.679 &   9.651 & 1.419 &  48631.787 &   9.555 & 1.405 &   50035.875 &   9.082 & 1.319\\
46679.979 &  9.287 & 1.375 &   47211.692 &   9.725 & 1.441 &  48635.787 &   9.566 & 1.392 &   50051.867 &   9.359 & 1.364\\
46703.992 &  9.351 & 1.373 &   47212.692 &   9.740 & 1.431 &  48636.708 &   9.562 & 1.394 &   50052.846 &   9.389 & 1.365\\
46704.979 &  9.351 & 1.377 &   47214.704 &   9.758 & 1.427 &  48638.812 &   9.555 & 1.391 &   50094.737 &   9.463 & 1.399\\
46716.963 &  9.421 & 1.392 &   47221.658 &   9.654 & 1.415 &  48645.787 &   9.478 & 1.381 &   50097.696 &   9.435 & 1.384\\
46720.954 &  9.403 & 1.389 &   47228.646 &   9.406 & 1.381 &  48646.779 &   9.459 & 1.375 &   50101.742 &   9.360 & 1.371\\
46723.963 &  9.364 & 1.386 &   47233.667 &   9.244 & 1.357 &  48649.767 &   9.411 & 1.389 &   50145.633 &   9.460 & 1.380\\
46724.950 &  9.343 & 1.402 &   47234.654 &   9.226 & 1.352 &  48652.754 &   9.375 & 1.378 &   50149.608 &   9.388 & 1.367\\
46728.958 &  9.238 & 1.367 &   47238.642 &   9.191 & 1.333 &  48693.642 &   9.398 & 1.389 &   50151.608 &   9.363 & 1.364\\
46731.950 &  9.149 & 1.354 &   47247.654 &   9.317 & 1.382 &  48697.633 &   9.420 & 1.373 &   50153.608 &   9.319 & 1.368\\
46739.921 &  9.061 & 1.337 &   47250.658 &   9.380 & 1.375 &  48866.975 &   9.259 & 1.358 &   50161.617 &   9.211 & 1.337\\
46740.971 &  9.057 & 1.357 &   47254.646 &   9.461 & 1.375 &  48874.967 &   9.088 & 1.355 &   50172.650 &   9.175 & 1.344\\
46741.946 &  9.074 & 1.338 &   47523.808 &   9.455 & 1.383 &  48888.925 &   9.408 & 1.382 &   50175.621 &   9.224 & 1.342\\
46742.954 &  9.062 & 1.360 &   47524.838 &   9.457 & 1.377 &  48891.904 &   9.466 & 1.366 &   50366.967 &   9.338 & 1.362\\
46747.933 &  9.104 & 1.335 &   47525.787 &   9.467 & 1.363 &  48927.879 &   9.396 & 1.340 &   50425.900 &   9.386 & 1.357\\
46759.933 &  9.326 & 1.328 &   47526.783 &   9.459 & 1.373 &  48976.812 &   9.384 & 1.355 &             &         &      \\
\hline
\end{tabular}
\end{center}
\end{table*}

AI~Aurigae was observed on 199 nights between 12 Oct 1985 and
8 Dec 1996. The observations were carried out at 
Lowell Observatory with the 0.53 m telescope. The detector
was a single-channel photoelectric photometer (EMI 6256S tube)
equipped with Str\"omgren $b$ and $y$ filters. The measurements
were obtained almost always through a 29$^{\prime\prime}$ diaphragm; some
measurements in bad seeing were made through a 49$^{\prime\prime}$ aperture,
and in bright Moonlight a 19$^{\prime\prime}$ aperture was occasionally used.

The differential magnitudes were aquired relative to HD~46159 ($V$=8\fm054,
$b-y$=0\fm623, G8III), while the photometric stability was checked against
HD~259662 ($V$=9\fm556, $b-y$=0\fm639, K0III).
The data were corrected only for differential extinction using
long-term monthly means for $k_b$ and $k_y$ (Lockwood \& Thompson 1986).
The standard transformations are represented with slight colour terms
(roughly $V=y+0.05(b-y)_i$, $b-y=1.08(b-y)_i$).
The internal consistency is of order of 0\fm005 as suggested by
the {\it rms} errors of the means for the comparison stars in $y$ and $b$
(0\fm0046 and 0\fm0048). The quoted standard values for the comparison
stars were determined on four nights
from standard star observations (extending to several very red
Landolt stars). Due to the red colour of the variable, there might
be some remaining systematic shift of 0\fm020-0\fm025 in the $V$
brightness. In any case, the scatter of the observations is
essentially negligible compared to the range of the variations.
The data (mean values of var$-$comp and var$-$check) are
presented in Table\ 1 (available electronically and/or in printed version?)
and plotted in Fig.\ 1.

% Fig. 1
\begin{figure}
\begin{center}
\leavevmode
\psfig{figure=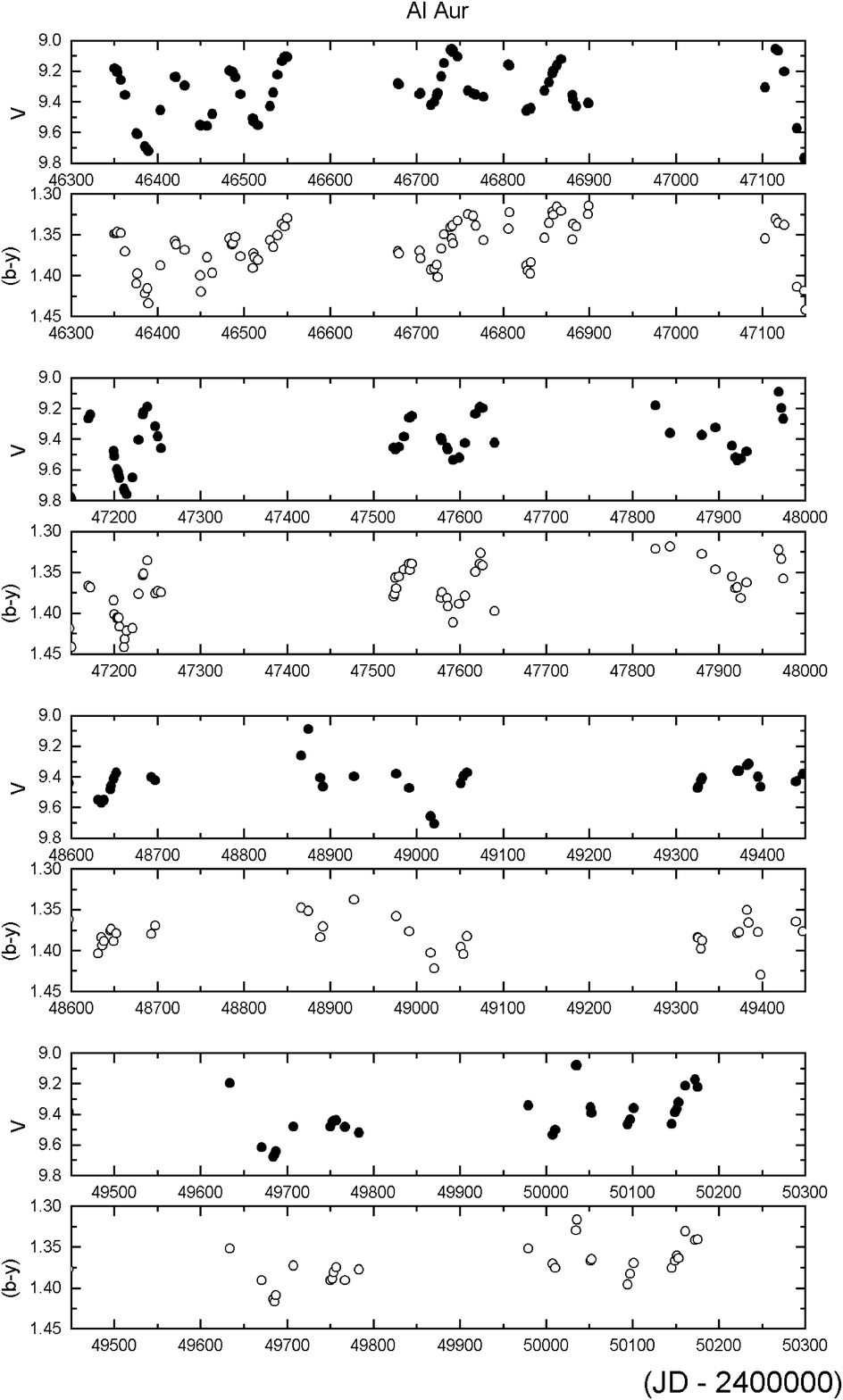,width=\linewidth}
\caption{The light and colour curves of AI~Aur between 1985 and 1996. Note the
600-day break between MJD 48000 and 48600}
\end{center}
\label{h2540f1}
\end{figure}

\section{Discussion}

In the following, we characterize the
observed light variations, discuss the importance of the colour
measurements and compare AI~Aur with other semiregulars with
available similar data series.

\subsection{Period analysis}

% Fig. 2
\begin{figure}
\begin{center}
\leavevmode
\psfig{figure=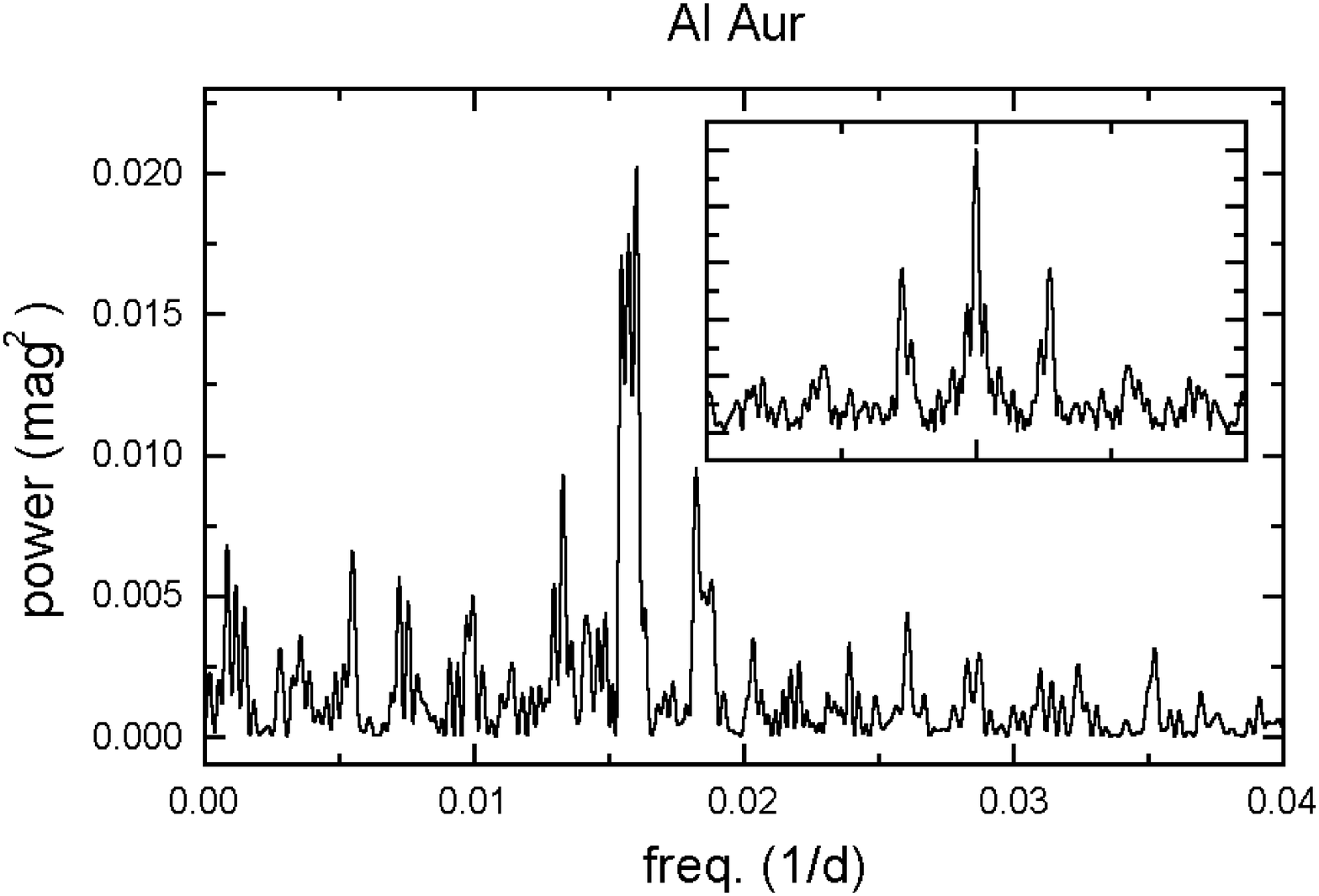,width=\linewidth}
\caption{The Fourier spectrum of the whole dataset. The insert
shows the window function with the same frequency scale}
\end{center}
\label{h2540f2}
\end{figure}

Data presented in Table\ 1 are plotted in Fig.\ 1. Note that some parts
are excluded from the presentation due to the sparse light curve coverage
(between MJD 48000-48600). However, when analysing the whole
dataset, we have also used those data that are not presented in Fig.\ 1.
Qualitatively the light curve is that of a typical semiregular without
strictly repeating cycles. This has been investigated in detail
by means of standard Fourier analysis implemented with Period98
of Sperl (1998).

First, we have calculated the power spectrum of the whole dataset.
It is shown in Fig.\ 2, where the insert shows the window function. One main
peak is present at f=0.01604 c/d (A=0\fm14) corresponding to
a period of 62\fd345. After a simple prewhitening with this frequency,
the power spectrum of the residuals contains a closely separated frequency
(f=0.01548 c/d, A=0\fm10).
However, it is a direct consequence of the unstable period, since
the whole dataset could be fitted neither with a single period nor
with a sum of two close periods. Therefore we studied the light curve
in four separate subsets (each being about 1000 days long) plotted
in the left panels of Fig.\ 3. We performed similar frequency analysis
as for the whole light curve and the results can be summarized as follows:

\begin{enumerate}

\item There is a well-defined main peak in every Fourier spectra
representing the average cycle length varying around the mean period of
$\sim$65 days. The fitted curves (solid lines in Fig.\ 3) have the
following periods and amplitudes: 1 - 63\fd2, 0\fm2; 2 - 67\fd1, 0\fm13;
3 - 100 d, 0\fm12 (the few data points may result in wrong result);
4 - 65 d, 0\fm15. The typical period uncertainty is about 1 d (except
for the third subset).

\item The first subset can be fitted very well with two
harmonic components, however the second one is simply due to the
variation of the mean brightness. Later data do not support the
presence of this false period ($\sim$860 days).

\item The phase diagrams of the individual subsets
(right panels of Fig.\ 3) suggest the average light curve shape
to be quite stable. The data were phased with those periods
listed above.

\item The prewhitened spectra of the subsets do not contain
any other significant peak higher than 0\fm08. We have not tried to
use the low-amplitude component(s), because the residual light curves
show irregular rather than cyclic behaviour. We feel that
when analysing light curves of SR variables the noise level in
the frequency spectrum is only partly due to the observational
errors and it would be misleading to accept every peaks in the
spectrum being much higher than the mean error of
the observations (about 0\fm01 in our case). The star itself
may grossly increase the noise of the light curve through its
erratic behaviour caused, e.g., by the mild chaos affecting
the pulsation (see e.g. Buchler \& Koll\'ath 2000). It is
well established, e.g. for Mira stars, that radial pulsation alone
may produce irregularities in the light and/or radial velocity
curves (for instance, through shock wave propagation in the extended
atmosphere, Bertschinger \& Chevalier 1985, Fox \& Wood 1985).
The complicated stellar responses on other physical mechanisms
(coupling between pulsation and convection, possible departures
from the spherical symmetry) prevent drawing a simple and
straightforward conclusion on the irregularities. Most recently,
Lebzelter et al. (2000) revived the old idea proposed first
by Schwarzschild (1975) on the large convective cells introducing
some irregularity in the light variation. However, in the
case of AI~Aur, only speculations can be drawn.

\item The results slightly depend on the selection
of the subsets (we have tried different lengths starting from one year
to the finally accepted 1000 days); on the other hand, they are
strongly affected by the lack of points in the second half of the data.

\end{enumerate}

% Fig. 3
\begin{figure*}
\begin{center}
\leavevmode
\psfig{figure=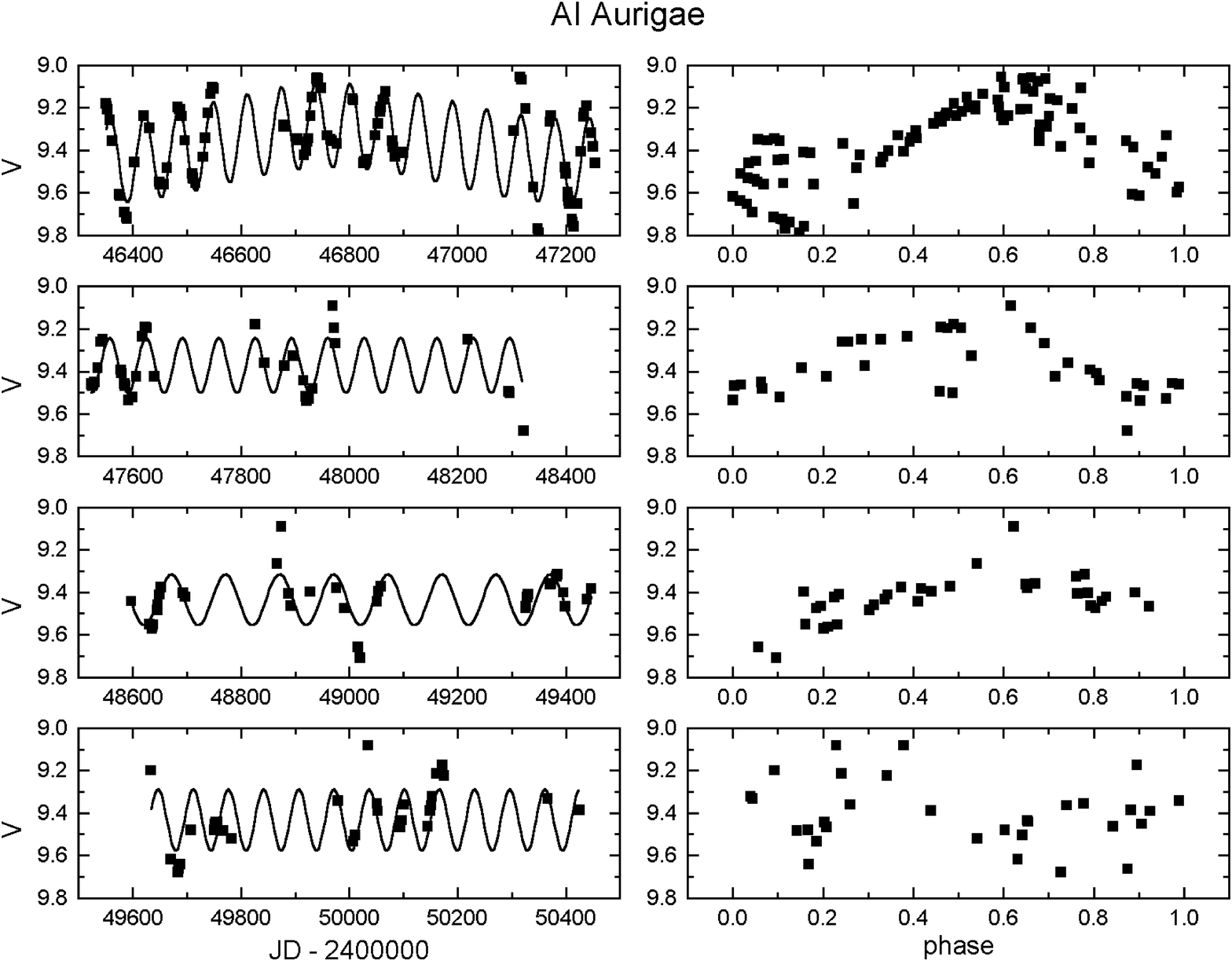,width=14cm}
\caption{The four subsets and the corresponding phase diagrams of
AI~Aur}
\end{center}
\label{h2540f3}
\end{figure*}

Considering these points we conclude that the light curve cannot be
described with simple sums of harmonic functions, but
it has a slowly and apparently irregularly variable characteristic
cycle length (instead of period) near 65 days.
The period analysis of the $b$-filter data leads to the
same conclusion.

\subsection{Colour variations}

The conclusions presented so far are typical for semiregular
variables (see, e.g. Kiss et al. 1999). The novelty in our results
is the presence of colour measurements. To our knowledge,
there have not been really
long-term (covering tens of cycles) time-series multicolour data
in the literature for any semiregular variable star.  Although Hipparcos
Epoch Photometry (ESA 1997) contains typically 100-200 points for thousands
of variables, the limited time span does not permit accurate period
determination for red variables.
High-precision V-band observations of different SRs were presented,
for instance, by Percy et al. 1989, Cristian et al. (1995), Percy et al.
(1996) and Lebzelter (1999), but none of these studies dealt with the colour
variations.

One of the most extensive time-resolved multicolour photometric
surveys was published by Smak (1964), who reported $UBV$ observations of
twenty-nine Mira stars and twelve semiregulars during two years, covering
typically only one cycle per star. He did not find any clear correlation
of the $B-V$ colour and the $V$ brightness for the SR stars.
Jerzykiewicz (1984) studied the light and colour variations
of HD~157010 (V818~Her), concluding that the star's $UBV$ magnitudes
are effective temperature parameters through the sensitivity of
the continuous spectrum and TiO blanketing. Wisse (1981) presented
$UBV$ observations for 35 SRs for classification purposes.
He found that well-defined correlations and anti-correlations
exist among the variables, where the phase difference of the
$V$ and $B-V$ maxima is either 0 or 0.5.
Cadmus et al. (1991), in their study of three SRs, noted only
that the colour changes were small compared to the changes in brightness.
Narrow-band observations of SR variables in the literature are
quite scanty. An interesting application can be found in
Wasatonic \& Guinan (1998), who used three colors of the Wing near-infrared
system to trace the temperature and radius variations of V~CVn.
These authors measured the highest temperature (from an
infrared index) around the maximum light.

The most important feature of our observations is the almost
perfectly parallel light and colour curves. This is a characteristic
behaviour in radially pulsating variables, where the temperature maximum
occurs close to the maximum light. It means that the star appears bluer when
it is at maximum light, which is in good agreement what is expected
from pure thermal variations driven by pulsation. In the case of FS~Comae
(SR star with P$\approx$58 d), Torres et al. (1993)
reached the opposite conclusion based on $UBV$ data series.
The star showed reversed colour changes explained
by the fact that $B-V$ colour in FS~Com is a molecular absorption index,
rather than a temperature indicator. The Str\"omgren $b$ filter
(centered near 4700\AA) has a relatively clear spectral region in the
center of the passband, but the wings on both sides have
moderately-weak TiO absorption, particularly the 2,0 transition
with bandhead at 4761\AA. The $y$ filter is not so strongly affected, with
only the usual Fe, Ti, etc. lines. Unlike the $B-V$ colour,
$b-y$ is completely unaffected by the wide TiO bandhead at
$\sim$5167\AA.
We conclude that the observations imply radial oscillation with dominant
thermal effects as the main reason for the light variability.
The $\sim$0\fm1 amplitude of the $b-y$ curve corresponds
to a temperature change of $<$300 K (as estimated from the
synthetic colour grids of Kurucz 1993).

\subsection{Stellar parameters}

Adopting radial pulsation for AI~Aur, further constraints can be drawn
on the basic stellar parameters. Unfortunately, there is no parallax
measurement for AI~Aur, thus only spectroscopic parallax or various
empirical period-colour-luminosity relations (e.g. Barth\`es et al. 1999)
can be used to determine
its luminosity. We have chosen the former approach, because
we wanted to estimate stellar properties without any assumption
on its pulsation. In the following we have neglected the
interstellar reddening since the observed mean $b-y$ colour (1\fm35)
is close to the expected value for an M5 giant star (see Jorissen
et al. 1995 for Str\"omgren photometry of red giants).

The IRAS [12]-[25] colour is $-1\fm41$. Following the
definitions by Kerschbaum \& Hron (1992), this means that AI~Aur
belongs to the ``blue'' semiregulars (no indication for circumstellar
shells, P$<$150$^d$, $T_{\rm eff}>3200$ K).
The infrared brightnesses ($K$=2\fm71, $I$=6\fm18),
taken from the IRC catalogue (Neugebauer \&
Leighton 1969), combined with the mean $V$ magnitude imply
colour indices $V-K$=6\fm8 and $V-I$=3\fm3.
The latter value results in a $T_{\rm eff}$=3420 K using the temperature
scale by Dumm \& Schild (1998). A less biased estimate can be inferred from
twenty M5III-type stars in the sample of Dumm \& Schild (1998):
$\langle T_{\rm eff}(M5III) \rangle $=3520$\pm$100 K.
For an independent check we have used the $(V-K)$-$(710-888)$ and
$(I-K)$-$(710-888)$ relations in Alvarez et al. (2000) resulting in
$T_{\rm eff}$=3300 K from their $T_{\rm eff}$-$(710-888)$ calibration.
Similarly, one can get a mean radius and mass for the given spectral
type of $\langle R \rangle = 123 \pm 34 R_\odot$ and
$\langle M \rangle = 2.0 \pm 0.8 M_\odot$. Adopting these
spectral type-mass and spectral type-radius values we calculated
a pulsational constant
$Q (=P \cdot \sqrt{ \rho / \rho_\odot})$=0.065$^{+0.06}_{-0.03}$.
Recently, Percy \& Parkes (1998) discussed the pulsation modes in
small-amplitude red variable stars reaching a conclusion that
some stars are likely to pulsate in up to the third overtone mode.
Within these frameworks AI~Aur seems to pulsate in the
fundamental or first overtone mode as is suggested by theoretical
models of Xiong et al. (1998) or Ostlie \& Cox (1986).
The overtone pulsation is favoured by the position of AI~Aur
in the $K$-band P-L diagram of long-period variables derived
by Bedding \& Zijlstra (1998) from Hipparcos parallaxes.
For this, we estimated $M_K$ from the $V-K$ colour and
$M_V$-spectral-type calibration of Th\'e et al. (1990).
The resulting $M_K=-7\fm5$ and log~P$\approx$1.8 place
AI~Aur close to the upper sequence in Fig.\ 1 of Bedding \& Zijlstra
(1998), which may be interpreted as a consequence of pulsation
in a different mode than that valid for most Mira stars
(represented by the lower sequence in Fig.\ 1 of Bedding \& Zijlstra
1998).

Finally, there is an interesting period-gravity relation for a wide
range of radially pulsating variable stars presented by Fernie (1995),
which can also be used to test the assumption of radial oscillation.
The adopted mass and radius give a log~$g$=0.56$\pm$0.4 which is
coincidentally the same as predicted by Eq.\ 1 of Fernie (1995).
Plotting AI~Aur in Fig.\ 1 of Fernie (1995), its position is as
deviant as that of Mira itself suggesting the first
overtone to be somewhat more likely. However, the universality of this
period-gravity relation has no firm theoretical background and
verification; therefore, this comparison should be considered only
as a possible hint for the mode of pulsation.

The presented considerations are on the whole
consistent with the recent observational and theoretical results
regarding the mode of pulsation in Mira and semiregular variables.
For example, Feast (1996) found that semiregular variables,
independently of their metallicity, pulsate probably in
the first overtone mode. Further supporting arguments were listed
by Feast (1999). However, semiregulars form a quite heterogeneous
group, in which
stars may pulsate in the fundamental, or 1st, 2nd or even 3rd
overtone mode, as has been clearly demonstrated by Wood et al. (1999).
That is why every individual case study has to be performed
without any definite preconception.

\section{Conclusions}

The main results presented in this paper can be summarized
as follows:

\noindent 1. New Str\"omgren $by$ photometric observations of the
semiregular variable AI~Aurigae are presented and discussed. The period
analysis of the whole light curve confirms the early period determination
of 63.9 days. However, the light curve shows typical semiregular
behaviour with changing amplitude and cycle length.

\noindent 2. The $b-y$ colour measurements revealed highly paralel
light and colour variations. Since the $b-y$ colour is considerably good
temperature indicator (rather than molecule indicator, as is
the $B-V$ colour), the phase dependent light and temperature
changes resemble classical radially pulsating stars. The observations
imply radial oscillation with dominant thermal effects as the main
reason for light variability.

\noindent 3. We estimated the basic stellar parameters, such as
the effective temperature (3520$\pm$100 K), mean radius (123$\pm$34
R$_\odot$), mass (2.0$\pm$0.8 M$_\odot$). The spectral type-mass
amd spectral type-radius values give a pulsational constant
of $0.065^{+0.06}_{-0.03}$. The results are consistent with
radial pulsation in the fundamental or (more likely) first
overtone mode. The period-gravity relation of Fernie (1995)
also supports this conclusion.

\begin{acknowledgements}

This research was supported by the ``Bolyai J\'anos'' Research
Scholarship of LLK from the Hungarian Academy of Sciences,
Hungarian OTKA Grant \#T032258 and Szeged Observatory Foundation.
The NASA ADS Abstract Service was used to access data and references.
This research has made use of the SIMBAD database, operated at CDS-Strasbourg,
France.

\end{acknowledgements}


\begin{thebibliography}{}

\bibitem[2000]{alvar2k}
    Alvarez R., Lan\c con A., Plez B., Wood P.R. 2000, A\&A 353, 322

\bibitem[1999]{barth99}
    Barth\`es D., Luri X., Alvarez R., Mennessier M.O. 1999, A\&AS 140, 55

\bibitem[1998]{bedding98}
    Bedding T.R., Zijlstra A.A. 1998, ApJ 506, L47

\bibitem[1985]{bertsch}
    Bertschinger E., Chevalier R.A. 1985, ApJ 299, 167

\bibitem[1937]{Beyer37}
    Beyer M., 1937, AN 262, 257

\bibitem[2000]{buchkoll}
    Buchler J.R., Koll\'ath Z. 2000, In: Takeuti M., Sasselov D. (eds.),
    Nonlinear Stellar Pulsation, Astrophysics and Space Science Library
    (AASL), in press

\bibitem[1991]{cadmus91}
    Cadmus R.R., Jr., Willson L.A., Sneden C., Mattei J.A. 1991,
    AJ 101, 1043

\bibitem[1973]{chuadze73}
    Chuadze A.D. 1973, Byull. Abastumanskaya Astrofiz. Obs. 44, 105

\bibitem[1995]{cris95}
    Cristian V.C., Donahue R.A., Soon W.H. et al. 1995, PASP 107, 411

\bibitem[1982]{dd82}
    Dickinson D.F., Dinger A.S.C 1982, ApJ 254, 136

\bibitem[1998]{dumm98}
    Dumm T., Schild H. 1998, New Ast. 3, 137

\bibitem[1997]{esa97}
    ESA 1997, The Hipparcos and Tycho Catalogues, ESA SP-1200

\bibitem[1996]{feast96}
    Feast M.W. 1996, MNRAS 278, 11

\bibitem[1999]{feast99}
    Feast M.W. 1999, IAU Symp. 191, 109

\bibitem[1972]{feast72}
    Feast M.W., Woolley R., Yilmaz N. 1972, MNRAS 158, 23

\bibitem[1995]{fernie95}
    Fernie J.D. 1995, AJ 110, 2361

\bibitem[1985]{fw85}
    Fox M.W., Wood P.R. 1985, ApJ 297, 455

\bibitem[1984]{jerz84}
    Jerzykiewicz M. 1984, AcA 34, 353

\bibitem[1995]{joriss95}
    Jorissen A., Mowlavi N., Sterken C., Manfroid J. 1995, A\&A 324, 578
    
\bibitem[1985]{gcvs}
    Kholopov P.N., Samus N.N., Frolov M.S., et al. 1985--88,
    General Catalogue of Variable Stars, $4^{th}$ edition,
    ``Nauka'' Publishing House, Moscow (GCVS4)

\bibitem[1992]{kersch92}
    Kerschbaum F., Hron J. 1992, A\&A 263, 97

\bibitem[1999]{kiss99}
    Kiss L.L., Szatm\'ary K., Cadmus Jr. R.R., Mattei J.A. 1999,
    A\&A 346, 542

\bibitem[2000]{kiss2k}
    Kiss L.L., Szatm\'ary K., Szab\'o Gy., Mattei J.A. 2000,
    A\&AS 145, 283

\bibitem[2000]{koen2000}
    Koen C., Laney D. 2000, MNRAS 311, 636

\bibitem[1993]{kur93}
    Kurucz R.L. 1993, ATLAS9 Stellar Atmosphere Programs and
    2 km/s Model Grids, CD-ROM No.13

\bibitem[1999]{lebz99a}
    Lebzelter T. 1999a, A\&A 346, 537

\bibitem[1999]{lebz99b}
    Lebzelter T. 1999b, A\&A 351, 644

\bibitem[1999]{letal99}
    Lebzelter T., Hinkle K.H., Hron J. 1999, A\&A 341, 224

\bibitem[2000]{letal2k}
    Lebzelter T., Kiss L.L., Hinkle K.H. 2000, A\&A 361, 167

\bibitem[1986]{lock86}
    Lockwood G.W., Thompson D.T. 1986, AJ 92, 976

\bibitem[1997]{lock97}
    Lockwood G.W., Skiff B.A., Radick R.R. 1997, ApJ 485, 789

\bibitem[1969]{irc69}
    Neugebauer G., Leighton R.B. 1969, Two micron sky survey:
    A preliminary Catalogue, Calif. Inst. Technology, NASA

\bibitem[1986]{oc86}
    Ostlie D.A., Cox A.N. 1986, ApJ 311, 864

\bibitem[1998]{percy98}
    Percy J.R., Parkes M. 1998, PASP 110, 1431

\bibitem[1989]{percy89}
    Percy J.R., Landis H.J., Milton R.E. 1989, PASP 101, 893

\bibitem[1996]{percy96}
    Percy J.R., Desjardnis A., Yu L., Landis H.J. 1996, PASP 108, 139

\bibitem[1964]{smak64}
    Smak J. 1964, ApJS 9, 141

\bibitem[1998]{sperl}
    Sperl M. 1998, Comm. Astr. Seis. 111

\bibitem[1990]{the90}
    Th\'e P.S., Thomas D., Christensen C.G., Westerlund B.E. 1990,
    PASP 102, 565

\bibitem[1993]{torres93}
    Torres G., Mazeh T., Latham D.W., Stefanik R.P. 1993, PASP 105, 360

\bibitem[1998]{vcvn98}
    Wasatonic R., Guinan E.F. 1998, IBVS No. 4579

\bibitem[1981]{wisse81}
    Wisse P.N.J. 1981, A\&AS 44, 273

\bibitem[1999]{wood99}
    Wood P.R., Alcock C., Allsman R.A. et al. (The MACHO Collaboration)
    1999, IAU Symp. 191, 151

\bibitem[1998]{xiong98}
    Xiong D.R., Deng L., Cheng Q.L. 1998, ApJ 499, 355

\end{thebibliography}
\end{document}